\documentclass[aip,apl,reprint,amsmath,amssymb]{revtex4-1}

\usepackage{graphicx}
\usepackage[utf8]{inputenc}
\usepackage[T1]{fontenc}
\usepackage{bm}
\usepackage{nicematrix}
\usepackage{color}
\usepackage{soul}

\usepackage{fancyhdr}
\usepackage{lipsum}

\begin{document}

\title{Nanodiamond quantum thermometry assisted with machine learning} %Title of paper

\author{Kouki Yamamoto}
\email{kouki.yamamoto@phys.s.u-tokyo.ac.jp}
\affiliation{Department of Physics, The University of Tokyo, Bunkyo-ku,Tokyo 113-0033, Japan}

\author{Kensuke Ogawa}
\affiliation{Department of Physics, The University of Tokyo, Bunkyo-ku,Tokyo 113-0033, Japan}
\author{Moeta Tsukamoto}
\affiliation{Department of Physics, The University of Tokyo, Bunkyo-ku,Tokyo 113-0033, Japan}
\author{Yuto Ashida}
\affiliation{Department of Physics, The University of Tokyo, Bunkyo-ku,Tokyo 113-0033, Japan}
\affiliation{Insititute for Physics of Intelligence, The University of Tokyo, Bunkyo-ku, Tokyo 113-0033, Japan}
\author{Kento Sasaki}
\affiliation{Department of Physics, The University of Tokyo, Bunkyo-ku,Tokyo 113-0033, Japan}
\author{Kensuke Kobayashi}
\affiliation{Department of Physics, The University of Tokyo, Bunkyo-ku,Tokyo 113-0033, Japan}
\affiliation{Insititute for Physics of Intelligence, The University of Tokyo, Bunkyo-ku, Tokyo 113-0033, Japan}
\affiliation{Trans-Scale Quantum Scinece Institute, The University of Tokyo, Bunkyo-ku, Tokyo 113-0033, Japan}

\date{\today}

\begin{abstract}
% insert abstract here
Nanodiamonds (NDs) are quantum sensors that enable local temperature measurements, taking advantage of their small size. Though the model based analysis methods have been used for ND quantum thermometry, their accuracy has yet to be thoroughly investigated. Here, we apply model-free machine learning with the Gaussian process regression (GPR) to ND quantum thermometry and compare its capabilities with the existing methods. We prove that GPR provides more robust results than them, even for a small number of data points and regardless of the data acquisition methods. This study extends the range of applications of ND quantum thermometry with machine learning.

\vspace{1em}
\begin{center}
\small
This is the Accepted Manuscript version of an article accepted for publication in \textit{Applied Physics Express}. IOP Publishing Ltd is not responsible for any errors or omissions in this version of the manuscript or any version derived from it. This Accepted Manuscript is published under a \href{https://creativecommons.org/licenses/by/4.0/}{CC BY licence}. The Version of Record is available online at \href{https://doi.org/10.35848/1882-0786/adac2a}{https://doi.org/10.35848/1882-0786/adac2a}.
\end{center}
\end{abstract}

\maketitle
\thispagestyle{fancy}

A diamond nitrogen-vacancy (NV) center is a point defect consisting of a nitrogen atom and an adjacent vacancy,  with an electron spin $S=1$. An NV center is an atom-sized sensor that can measure local physical quantities such as temperature~\cite{Acosta:temperature,Chen:temperature,Kucsko:thermometry,Temp:Neumann,Doherty:temperature,FujiwaraPRReseach,FujiwaraSciAdv,Petrini2022,Okabe2018,Temp:Liu,Ogawa:nanodiamond}, magnetic field~\cite{Mag:Hong,Mag:Barry,Mag:Mamin,Mag:Grinolds,Mag:McGuinness}, electric field~\cite{Ele:Dolde}, and pressure~\cite{pressure:Doherty,pressure:Ivady,pressure:Hsieh}. In particular, thermometry using NV centers is expected to be valuable in various fields, from life sciences to condensed matter physics. For example, the temperature change was observed inside cells or microorganisms using nanodiamonds (NDs) injected into them~\cite{Kucsko:thermometry,FujiwaraSciAdv,Petrini2022,Okabe2018}, and the thermal diffusion from heat sources was detected with NDs spread over the target materials~\cite{Temp:Neumann,Temp:Liu,Ogawa:nanodiamond}. Thus, local thermometry by NV centers provides a unique opportunity in nanoscience and nanotechnology.

Electron spin resonance in NV centers can be detected by measuring photoluminescence (PL) intensity while irradiating lasers and microwaves, so-called optically detected magnetic resonance (ODMR)~\cite{Mag:Barry}. We can measure temperature by estimating the temperature-dependent zero-field splitting (ZFS)~\cite{Cambria2023} of the NV center from the ODMR spectra. It is necessary to be careful when measuring temperatures using NV centers. Bulk diamond is invasive when measuring temperature distribution because of its high thermal conductivity~\cite{Tanos:conductivity}. In contrast, the ensemble of NDs, $5$\,--\,$200\,\mathrm{nm}$-sized diamond crystals~\cite{Schirhagl:nanodiamond}, have less thermal conductivity than bulk diamonds~\cite{Tanos:conductivity}. Therefore, NDs are more suitable for local thermometry.

The accuracy of the analysis for ND quantum thermometry needs to be clarified. It is difficult to accurately analyze the ODMR spectra of NDs since the crystal orientation of NDs is distributed in various directions~\cite{foy2020}. The 4-point method~\cite{Kucsko:thermometry,FujiwaraPRReseach}, which measures only four frequencies in the ODMR spectrum, is successfully applied for fast thermometry. However, the accuracy in this case must be carefully considered. Also, the fitting method, which fits the ODMR spectra by a double Lorentzian function, has been widely used, but the fits need to be more consistent~\cite{FujiwaraPRReseach,Tsukamoto:ML}. In addition, in a magnetic field, the spectra spread in a complex manner, making the analysis even more challenging with these methods~\cite{foy2020}.

In this work, we investigate the applicability of Gaussian process regression (GPR)~\cite{GPR:mackay,GPR:rasmussen}, a machine learning method, for ND quantum thermometry. We apply the conventional 4-point and fitting methods and the GPR method to analyze the spectra. To examine the robustness of the data analysis, we investigate each method's dependence on the number and selection of analysis data points in the ODMR spectrum. We show that the GPR method provides more robustly accurate thermometry than the existing methods.

\begin{figure}[htb]
\centering
\includegraphics[width=\linewidth]{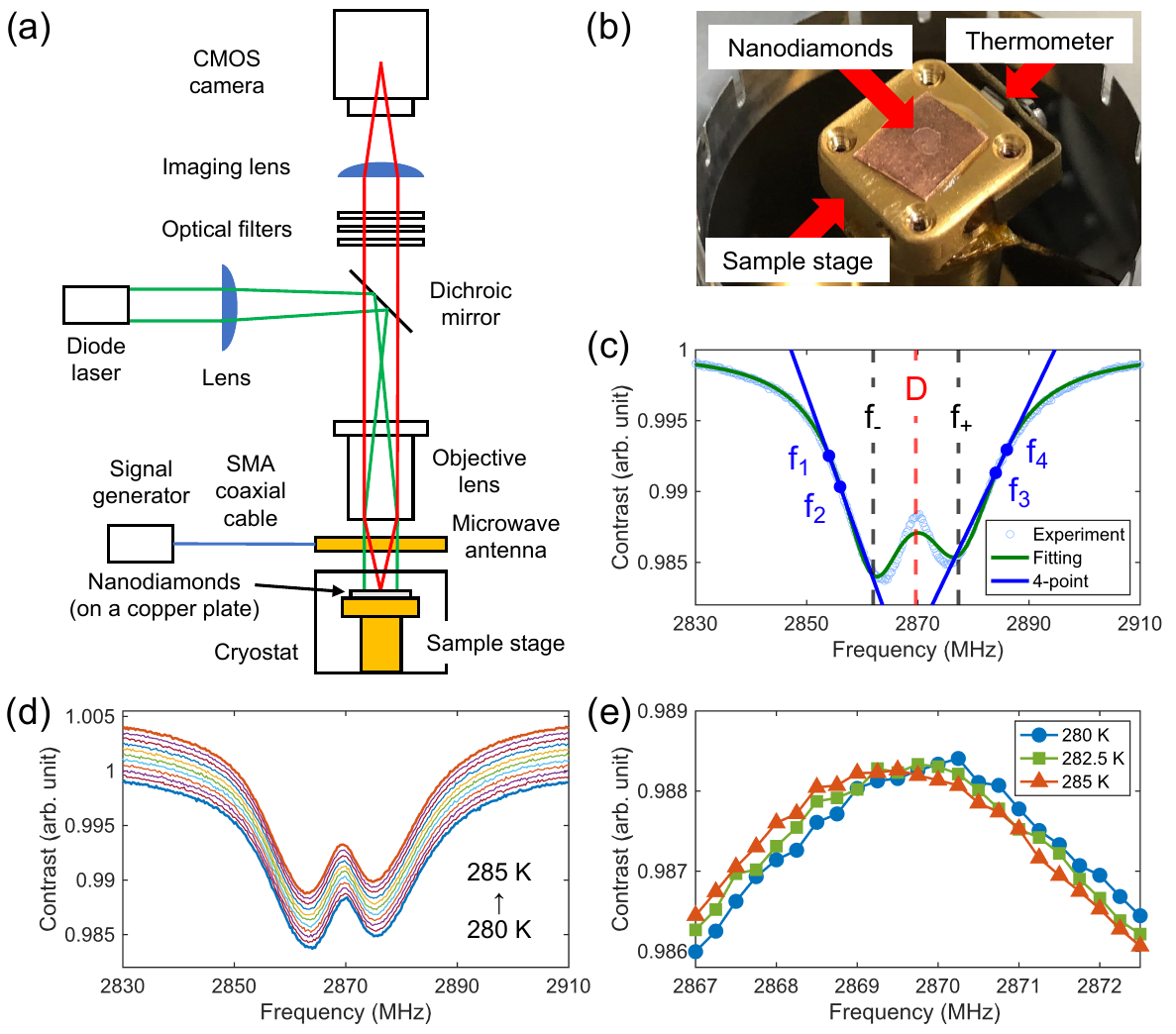}
\caption{(a) Overview of the experimental setup. (b) A view of the sample stage. NDs are spread on the copper plate. (c) Overview of the 4-point and fitting methods. The light blue open circles represent experimental data. The green line is fitted results using a double Lorentzian function. The four filled blue circles and blue lines represent the 4-point method. The black and red dashed lines respectively represent the resonance frequencies $f_\pm$ and the ZFS value $D$ obtained by fitting the ODMR spectrum with a double Lorentzian. (d) ODMR spectra from 280\,K to 285 \,K in increments of 0.5\,K. The bottommost and topmost lines represent spectra at 280\,K and 285\,K, respectively. Each spectrum has been incrementally shifted vertically for the sake of visibility. (e) An enlarged figure around the center (2870\,MHz) of ODMR spectra at 280\,K, 282.5\,K, and 285\,K in Fig.\,\ref{f1}(d).}
\label{f1}
\end{figure}

We first acquire a complete dataset of ODMR spectra in a precisely temperature-controlled setup, and then the results are analyzed using three methods, namely the 4-point, fitting, and GPR methods.
Figure \ref{f1}(a) shows our setup. 
The NDs are of $100~\,\mathrm{nm}$ particle size. NDs are spread thinly on a $100~\,\mathrm{\mu m}$ thick copper plate~\cite{Ogawa:nanodiamond}. 
The copper plate is placed on a sample stage in a cryostat with temperature stability better than $10~\,\mathrm{mK}$ [Fig.\,\ref{f1}(b)]. 
The copper plate with NDs and the sample stage are in sufficient thermal contact using thermal grease.
We assume that the temperatures of NDs are the same as the value of the calibrated thermometer embedded in the sample stage. 
We define this value as the ``true'' temperature $T_\mathrm{true}$.
An optical window is mounted on the top surface of the cryostat directly above the sample stage, and emissions from the NDs are measured through this window. 
A resonator microwave antenna~\cite{Sasaki2016} is placed directly above the optical window, and magnetic resonance of the NDs is performed~\cite{Nishimura2023}.

We obtain ODMR spectra of NDs at 11 different temperatures from 280~K to 285~K in 0.5~K increments (i.e., temperature points $N=11$).
When acquiring the ODMR spectra, the microwave frequency is swept at 321 equally spaced points from 2830~MHz to 2910~MHz. 
We randomly select the frequency points for each sweep to suppress frequency-dependent heating due to microwave antenna characteristics~\cite{Nishimura2023}. We adopt the result of integrating the PL intensity for 100 sweeps as a single ODMR spectrum data.
A typical result of a single ODMR spectrum is shown in Fig.~\ref{f1}(c). Two of the ODMR spectrum data are acquired at each temperature. In the following, the first data will be referred to as Data \#1 and the second as Data \#2. These datasets will be used later in the accuracy evaluation.

Figure~\ref{f1}(d) shows the ODMR spectra at all temperature conditions, and Fig.~\ref{f1}(e) is a magnified view of the data at 280~K, 282.5~K, and 285~K.
The center of the ODMR spectrum shifts to lower frequencies with increasing temperature.

The shift corresponds to a change in ZFS with temperature~\cite{Acosta:temperature,Chen:temperature}. The  4-point and fitting methods use a linear relationship between ZFS and temperature. For example, near room temperature, the ZFS value $D$ depends linearly on temperature $T$ with a proportionality coefficient $\alpha \equiv dD/dT \sim -74\,\mathrm{kHz/K}$~\cite{Acosta:temperature}.
Once the ZFS at a single reference temperature $T_0$ and a coefficient $\alpha$ are calibrated, the ZFS at a specific temperature $T$ can be shown as the following:
\begin{eqnarray}
    D(T) = \alpha(T - T_0) + D(T_0).\label{calc_temp}
\end{eqnarray}
The temperature can be obtained directly from the ZFS based on Eq.~(\ref{calc_temp}). Note that the coefficient $\alpha$ varies depending on the temperature region~\cite{Chen:temperature}. Hence, to circumvent the disadvantages of the conventional methods assuming the linearity, experiments are performed over a narrow temperature range, from 280\,K to 285\,K.

On the other hand, the GPR method estimates temperature from ODMR spectra by learning the relationship between ODMR spectra and temperature, including ZFS information (see ``Gaussian process regression (GPR)'' in ``Supplementary material'' for details). In the previous study, we used the GPR method to estimate the magnetic field accurately from ODMR spectra showing complex shapes in a magnetic field~\cite{Tsukamoto:ML}. The present study aims to verify how helpful the idea of magnetic field estimation using GPR is for temperature estimation.

\begin{figure}[b]
\centering
\includegraphics[width=\linewidth]{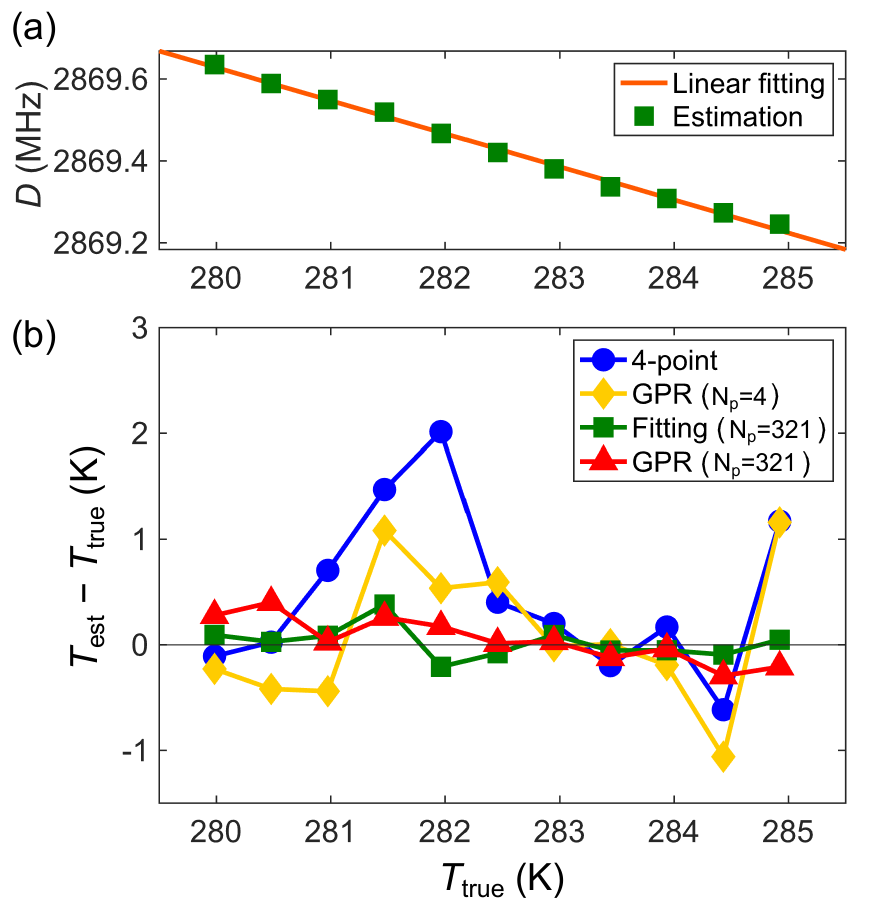}
\caption{(a) Observed ZFS value $D$ as a function of the ``true'' temperature $T_\mathrm{true}$. (b) Comparison of temperature estimation by the 4-point method, the fitting method ($N_p=321$), and the GPR methods ($N_p=4$ and $N_p=321$).
}
\label{f2}
\end{figure}

We rely on Data \#1 taken at a given $T_\mathrm{true}$ and estimate the temperature $T_\mathrm{est}$ from Data \#2 at each temperature.
We adopt the three methods mentioned above to obtain the $T_\mathrm{est}$ and compare their estimation accuracy. We use the following root-mean-square error (RMSE) calculated over 11 measured temperature points ($N=11$) as a measure of the appropriateness of the estimation:
\begin{equation}
    \mathrm{RMSE} = \sqrt{\frac{1}{N}\sum_{i=1}^N(T_\mathrm{est}^i-T_\mathrm{true}^i)^2},
\end{equation}
where $T_\mathrm{est}^i$ and $T_\mathrm{true}^i$ are estimated and true temperatures at the $i$-th set temperature ($i\in\{1, 2, \cdots, 11\}$), respectively. We use Data \#1 as calibration data for the 4-point and fitting methods and training data for the GPR method. Data \#2 is used as test data to obtain RMSE (see "Evaluation of estimation accuracy" in "Supplementary material" for details).

The ZFS estimated by fitting a double Lorentzian to all the ODMR spectra in Fig.\,\ref{f1}(d) is shown in Fig.\,\ref{f2}(a).
The linearity of ZFS is well maintained over the temperature range tested in this study. A linear fit provides the coefficient of $\alpha = -80.8\pm 1.8\,\mathrm{kHz/K}$ (the error is a fitting error, and the error range is set to $1\sigma$). It is consistent with the literature values ~\cite{Acosta:temperature,FujiwaraPRReseach,foy2020}.

Achieving accuracy with a few experimental data points is essential for a practical purpose. To pursue this, the accuracy of each method is also examined by varying the number of analysis data points $N_p$. 
When changing $N_p$, we adopt equally separated frequency points from the entire measurement frequency range except for $N_p=4$. For $N_p=4$ only, the same four points used in the 4-point method are adopted, and their frequencies are 2854~MHz, 2856~MHz, 2884~MHz, and 2886~MHz unless otherwise stated.

Figure\,\ref{f2}(b) shows the discrepancy between $T_\mathrm{est}$ and $T_\mathrm{true}$. The reference temperature to deduce $T_\mathrm{est}$ using Data \#2 in the 4-point method is 279.89\,K. The results of the 4-point method show sharp bumps when the temperature is varied. 
The GPR ($N_p=4$) result also shows bumps but to a smaller extent than the 4-point method. On the other hand, the temperature estimations by the fitting ($N_p=321$) and GPR ($N_p=321$) methods provide consistent results with $T_\mathrm{true}$. 
Regarding robustness, the fitting ($N_p=321$) and GPR ($N_p=321$) methods outperform the 4-point method, which is unsurprising as $N_p$ is large.
Nevertheless, the present result implies that GPR has the advantage even for $N_p=4$. 
We will show later that this is not a coincidence.

\begin{figure}[t]
\centering
\includegraphics[width=\linewidth]{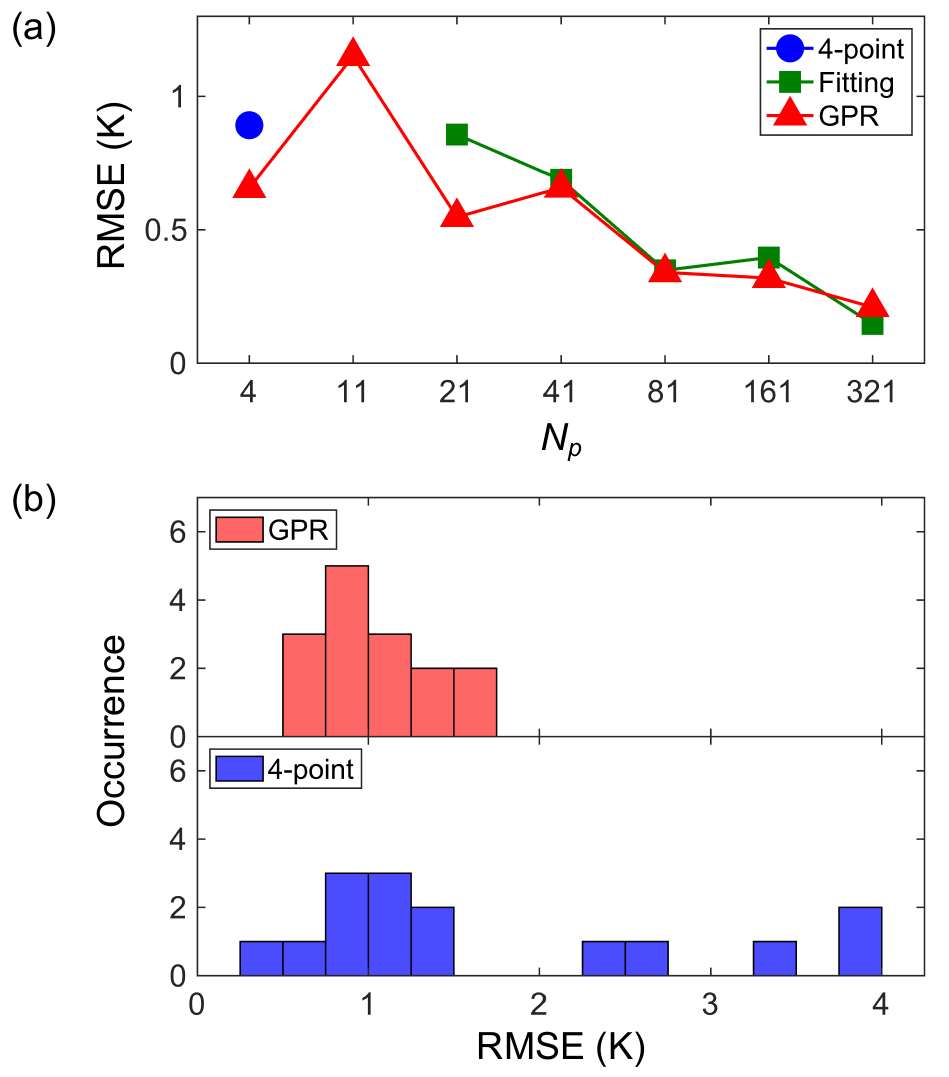}
\caption{(a) Comparison of RMSEs for each method when $N_p$ is varied. (b) Histograms of RMSEs calculated for the results of the GPR and 4-point methods in the top and bottom panels, respectively.
}
\label{f3}
\end{figure}

More quantitatively, Fig.~\ref{f3}(a) compares the RMSEs of each method when $N_p$ is varied. The fitting method does not work appropriately for $N_p=4$ and $N_p=11$, so they are not shown here. The result  indicates that the GPR method can robustly estimate temperatures from small $N_p$ to large $N_p$. The larger the $N_p$ is, the smaller the RMSE is. The GPR method allows analysis even with a small $N_p$ and uses the data efficiently to improve accuracy when a large $N_p$ is available. This fact meets our naive demand for a robust analysis method. On the other hand, the fitting method requires a sufficient $N_p$ for estimation. Moreover, it has become apparent during the analysis that the RMSE of the fitting method sensitively depends on the initial value of the fitting parameters.

\begin{table}[t]
    \centering
    \begin{NiceTabular}{c|c|c|c}
    \hline
    \Block{2-1}{Index} & \Block{2-1}{Frequencies $-2870$ (MHz)} & \Block{1-2}{RMSE (K)}\\\cline{3-4}
     & & GPR & 4-point\\
    \hline\hline
    1 & $-$20, $-$18, 18, 20 & 1.5960 & 3.8031 \\
    2 & $-$20, $-$16, 16, 20 & 1.2137 & 3.4996 \\
    3 & $-$20, $-$14, 14, 20 & 1.0409 & 2.3711 \\
    4 & $-$20, $-$12, 12, 20 & 0.8251 & 2.5814 \\
    5 & $-$20, $-$10, 10, 20 & 1.3989 & 3.8989 \\
    6 & $-$18, $-$16, 16, 18 & 1.5120 & 1.0629 \\
    7 & $-$18, $-$14, 14, 18 & 0.8915 & 0.3540 \\
    8 & $-$18, $-$12, 12, 18 & 0.8109 & 0.6848 \\
    9 & $-$18, $-$10, 10, 18 & 1.3751 & 1.3421 \\\hline
    $\bm{10}$ & $\bm{-16, -14, 14, 16}$ & $\bm{0.6561}$ & $\bm{0.8920}$ \\\hline
    11 & $-$16, $-$12, 12, 16 & 0.7113 & 0.9174 \\
    12 & $-$16, $-$10, 10, 16 & 1.2496 & 1.3424 \\
    13 & $-$14, $-$12, 12, 14 & 0.7879 & 0.9265 \\
    14 & $-$14, $-$10, 10, 14 & 0.9715 & 1.0014 \\
    15 & $-$12, $-$10, 10, 12 & 0.7055 & 1.1040 \\
    \hline
    \end{NiceTabular}
    \caption{Fifteen patterns of four frequency points used in the analysis of Fig.\,\ref{f3}(b). 
    The frequencies indicate the difference from 2870 MHz.
    Index 10 (written in bold) is used for the 4-point method result shown in Figs.~\ref{f2}(b) and \ref{f3}(a).
    }
    \label{t1}
\end{table}

It is interesting to compare the 4-point and GPR methods in more detail. The analysis method with fewer data points is more advantageous experimentally. However, the 4-point method is arbitrary in how the four points are taken, and the accuracy may depend on the way. Four particular points in the slope regions of the spectrum are required for the 4-point method, and there are many options for their selection. Thus, we select four points (two points each on the high-frequency and low-frequency side) in the region with slopes 10\,MHz to 20\,MHz away from 2870~MHz. Setting the frequency resolution at 2~MHz, selectable four points are restricted to 15 patterns, as listed in Table~\ref{t1}. RMSEs are calculated for the 15 patterns (see Table~\ref{t1}). In the 4-point method, the proportionality coefficient $\alpha$ in Eq.\,(\ref{calc_temp}) is calculated from Data \#1 and used for each of the 15 different four points when determining the RMSE.

Histograms of the 15 calculated RMSEs for the 4-point and GPR methods are shown in Fig.~\ref{f3}(b). The RMSE values are distributed over a narrow range for the GPR method, while they are spread over a wide range for the 4-point method. It suggests that the accuracy of the 4-point method is sensitive to the selection of the four points. In contrast, the GPR method is robust and less dependent on the data point selection. Therefore, even when $N_p=4$, the GPR method is more accurate and robust overall than the 4-point method.

The RMSE of the 4-point method in Index\,7 of Table\,\ref{t1} is minimal. Notably, the worth of the 4-point method can be increased if we clarify how to choose the four points improving its accuracy. However, such a method has yet to be found.

Note that Data \#1 and \#2 were acquired under the same conditions. The previous study~\cite{Tsukamoto:ML} on magnetic field measurements has shown the effectiveness of GPR, even for data acquired under different conditions. However, further study is needed to determine the effectiveness of GPR-based ND quantum thermometry under different conditions.

To conclude, we have established an ND quantum thermometry using GPR and shown that the present GPR method is an excellent analytical method with superior accuracy and robustness compared to the existing methods.

Developing a multi-output GPR method~\cite{GPR:rasmussen} that combines the accurate magnetic field estimation shown in the previous study~\cite{Tsukamoto:ML} with the accurate thermometry in the present study enables simultaneous accurate measurement of local temperature and magnetic field. For example, such a method is promising  in investigating the various phase transitions, including current-induced ones~\cite{simultaneous:Healey}. 
Incorporating ND quantum thermometry and magnetometry will expand the range of measurements of nanoscale physical properties.

\vskip\baselineskip

This work was partially supported by
%%% list of acknowledgments
JST, CREST Grant No.~JPMJCR23I2, Japan; 
Grants-in-Aid for Scientific Research (Nos.~JP22K03524, JP23H01103, JP22KJ1058, and JP22KJ1059); 
the Mitsubishi Foundation (Grant No.~202310021);
Kondo Memorial Foundation; Daikin Industry Ltd; 
the Cooperative Research Project of RIEC, Tohoku University;
``Advanced Research Infrastructure for Materials and Nanotechnology in Japan (ARIM)'' (No.~JPMXP1222UT1131) of the Ministry of Education, Culture, Sports, Science and Technology of Japan (MEXT). 
Y.A. acknowledges support from the Japan Society for the Promotion of Science (JSPS) through Grant No. JP19K23424 and from JST FOREST Program (Grant No. JPMJFR222U, Japan). 
K.O. and M.T. acknowledge supports from FoPM, WINGS Program, The University of Tokyo.
K.Y. acknowledges supports from JST SPRING Grant No. JPMJSP2108, Japan, and MERIT-WINGS, The University of Tokyo.

\bibliography{reference}

\end{document}